\documentclass{article}
\usepackage[utf8]{inputenc}
\usepackage{graphicx,xcolor,setspace,geometry}
\usepackage{amsmath,amssymb}
\usepackage{natbib,authblk,bbm,bm}
\usepackage{multirow}
\usepackage{booktabs}

\usepackage{tikz}
\usetikzlibrary{"automata","positioning"}
\usetikzlibrary{fit}
\usetikzlibrary{calc}
\usetikzlibrary{decorations.pathreplacing,tikzmark}
\usetikzlibrary{arrows.meta}
\tikzset{>={Latex[width=1.5mm,length=1.5mm]}}

\title{Flexible estimation of the state dwell-time
distribution in hidden semi-Markov models}
\author{
Jennifer Pohle$^{1}$\footnote{
Corresponding author; email: \texttt{jennifer.pohle@uni-bielefeld.de}; postal address: Universit\"atsstraße 25, 33615 Bielefeld, Germany.
}, 
Timo Adam$^{2}$, and 
Larissa T. Beumer$^{3}$\\
$^1$Bielefeld University, Germany\\
$^2$University of St Andrews, UK\\
$^3$Aarhus University, Denmark
}
\date{}

\begin{document}

\begin{spacing}{1.5}
\maketitle
\vspace{-2em}

\begin{abstract}
Hidden semi-Markov models generalise hidden Markov models by explicitly modelling the time spent in a given state, the so-called dwell time, using some distribution defined on the natural numbers. While the (shifted) Poisson and negative binomial distribution provide natural choices for such distributions, in practice, parametric distributions can lack the flexibility to adequately model the dwell times. To overcome this problem, a penalised maximum likelihood approach is proposed that allows for a flexible and data-driven estimation of the dwell-time distributions without the need to make any distributional assumption. This approach is suitable for direct modelling purposes or as an exploratory tool to investigate the latent state dynamics. The feasibility and potential of the suggested approach is illustrated by modelling muskox movements in northeast Greenland using GPS tracking data. The proposed method is implemented in the \texttt{R}-package \texttt{PHSMM} which is available on CRAN.

\end{abstract}
\vspace{0.5em}

\section{Introduction}

Hidden Markov models (HMMs) are flexible probabilistic models for sequential data which assume the observations to depend on an underlying latent state process.
Emerging from the field of speech recognition \citep{rab89}, they find applications in various areas, such as medicine \citep{lan13}, psychology \citep{vis02}, finance \citep{ngu18}, and ecology \citep{beu20}, where they are used for classification tasks, forecasting, or general inference on the data-generating process; for an overview of the various HMM applications, see, for example, \citet{zuc16}. In an HMM's basic model formulation, the underlying state sequence is assumed to be a finite-state first-order Markov chain. This assumption is mathematically and computationally very convenient and allows for an efficient likelihood evaluation and inference \citep{zuc16}. However, it also implicitly restricts the state dwell time, that is the number of consecutive time points spent in a given state, to follow a geometric distribution. Thus, the modal dwell time is fixed at one and the dwell time's distributional shape, with a strictly monotonically decreasing probability function, is completely predefined \citep{lan11}. This might be appropriate for some applications, but inappropriate or too restrictive for others. Examples for the latter include the modelling of daily share returns \citep{bul06}, the analysis of rainfall event data \citep{san01}, and speech unit modelling \citep{gue90}.

Hidden semi-Markov models (HSMMs) overcome this limitation by assuming the underlying state sequence to be a semi-Markov chain, thereby allowing for arbitrary dwell-time distributions defined on the natural numbers. First introduced in the field of speech recognition \citep{fer80}, the additional flexibility makes HSMMs attractive for various areas of application; an overview is provided by \citet{yu10}. However, in order to formulate an HSMM and apply it to data, again some class of dwell-time distributions must be chosen. This raises a new problem: How to select distributions which adequately describe the unobserved states' dwell times? The usual choice is a family of standard discrete parametric distributions, such as the (shifted) Poisson or negative binomial \citep{bul06,eco14,van15}. In that case, the geometric dwell-time distribution implied by conventional HMMs is replaced by another parametric distribution, which again corresponds to a restrictive assumption on the distribution's shape, and hence on the way the state process evolves over time.

An alternative approach which avoids restrictions on the distribution's shape is the use of discrete non-parametric distributions, that is, for each dwell time and state, an individual dwell-time probability is estimated (see, for example, \citealp{san01, gue03}). Such procedures usually require finite dwell-time domains with fixed maximum dwell times for each state \citep{bul10}. This is not necessarily restrictive if the domain is chosen large enough to capture the main dwell-time support, however, a large domain implies a large number of parameters to be estimated. Thus, usually, a large number of observations is needed to fit the model \citep{bul10}. More importantly, there is a high risk to obtain wiggly dwell-time distributions with implausible gaps and spikes. Consequently, the estimation could suffer from both overfitting and numerical instability due to probabilities estimated close to zero.

We aim to overcome these problems by proposing a penalised maximum likelihood (PML) approach that allows for the exploration of the underlying state dynamics in a data-driven way while providing flexible yet smooth estimates. Our method is built on dwell-time distributions with an unstructured (i.e.\ `non-parametric') start and a geometric tail \citep{san01, lan11} to avoid the use of finite dwell-time domains. The introduced penalty term then penalises higher-order differences between adjacent dwell-time probabilities of the unstructured start. This leads to smoothed probability functions and thereby helps to avoid overfitting. Using a state expansion trick, the considered HSMM can exactly be represented by an HMM, thereby opening the way for an efficient likelihood evaluation and numerical (penalised) maximum likelihood estimation \citep{lan11}. 
The remaining paper is structured as follows: In Section \ref{Sec2}, we discuss the HSMM model formulation and introduce our PML approach. Section \ref{Sec3} illustrates the feasibility and potential usefulness of the method with a real data case study using movement data from a muskox tracked in northeast Greenland. We conclude with a discussion in Section \ref{Sec4}.

\section{Methodology}\label{Sec2}

\subsection{Hidden semi-Markov models}\label{Sec2.1}

An HSMM is a doubly stochastic process comprising a latent $N$-state semi-Markov chain $\{S_{t}\}_{t=1}^{T}$ and an observed state-dependent process $\{Y_{t}\}_{t=1}^{T}$. Its basic dependence structure is illustrated in Figure \ref{fig:HSMM}. The model assumes that at each time point, the observation $Y_{t}$ is generated by one out of $N$ \textit{state-dependent distributions} $f(y_{t}|S_{t}=i)=f_{i}(y_{t})$, $i=1,\ldots,N$, as selected by the current state. Thus, given the current state $S_t=s_t$, $Y_{t}$ is assumed to be conditionally independent of past observations and states. Note that here, $f$ is used either to denote a probability mass function, if $Y_{t}$ is discrete, or a density function, if $Y_{t}$ is continuous-valued. For multivariate time series, $\mathbf{Y}_{t}=(Y_{1,t},\ldots,Y_{p,t})$, another simplifying assumption is often made, that is, given the current state $S_{t}=s_{t}$, the observations are contemporaneously conditionally independent of each other: $f(\mathbf{y}_t|S_{t}=s_{t})=\prod_{k=1}^{p} f(y_{k,t}|S_{t}=s_{t})$. This allows to choose suitable classes of univariate distributions for the different variables observed. Alternatively, multivariate state-dependent distributions can be used.

The underlying semi-Markov chain $\{S_{t}\}_{t=1}^{T}$ is described by two components: (i) Whenever the chain enters a new state $i$ at some time point $t$, a draw from the corresponding \textit{state dwell-time distribution} $d_{i}$ determines the number of consecutive time points the chain spends in that state. It is defined by its probability mass function (PMF)
$$
d_{i}(r)=\Pr(S_{t+r} \neq i, S_{t+r-1}=i,\ldots,S_{t}=i|S_{t}=i,S_{t-1} \neq i),
$$
with $r\in \mathbb{N}$ denoting the duration; (ii) The state switching is described by an \textit{embedded Markov chain} with \textit{conditional transition probabilities} $\omega_{ij}=\Pr(S_{t}=j|S_{t-1}=i, S_{t} \neq i)$, summarised in the $N \times N$ conditional transition probability matrix $\boldsymbol{\Omega}$ with $\omega_{ii}=0$. The \textit{initial distribution} describes the state probabilities at $t=1$, $\bm{\delta}=(\Pr(S_{1}=1),\ldots,\Pr(S_{1}=N))$.
 
In case that all state dwell times are geometrically distributed, the HSMM reduces to the special case of an HMM and the underlying state-sequence $\{S_{t}\}_{t=1}^{T}$ becomes a first-order Markov chain. The state-switching is then characterised by the $N \times N$ \textit{transition probability matrix} (TPM) $\boldsymbol{\Gamma}=(\gamma_{ij})$ with $\gamma_{ij}=\Pr(S_{t}=j|S_{t-1}=i)$ denoting the \textit{transition probabilities}. This automatically implies the geometric dwell-time distribution with $d_{i}(r)=(1-\gamma_{ii})\gamma_{ii}^{r-1}$ for each state $i=1,\ldots,N$.

The parameter vector $\bm{\theta}$ characterising an $N$-state HSMM contains the parameters defining the dwell-time distributions $d_i(r)$ and the state-dependent distributions $f_{i}(y_{t})$, for $i=1,\ldots,N$, the conditional transition probabilities $\omega_{ij}$, for $i,j=1,\ldots,N$, $i\neq j$, and the initial probabilities $\delta_i$, $i=1,\ldots,N$. Thus, for parameter estimation, it is necessary to choose classes of parametric or non-parametric state-dependent and state dwell-time distributions. Although not trivial, the former can usually be chosen and evaluated directly based on an inspection of the observations at hand. For instance, for daily share return data, normal or t-distributions are common options \citep{bul06, oel20}, and for movement data, gamma or Weibull distributions are often suitable to model the observed step lengths \citep{lan12}. The state dwell times, however, are usually unobserved, which makes the choice of appropriate distributions difficult. As a way to solve this problem, in the subsequent section, we propose a penalised maximum likelihood approach which avoids strong assumptions about the distributions' shape.

\begin{center}
\begin{figure}[!t]
	\centering
	\begin{tikzpicture}[node distance = 1.5cm]
	\tikzset{state/.style = {circle, draw, minimum size = 55pt, scale = 0.725}}
	\node [state] (1) at (0,0) {$S_{t-1}=j$};
	\node [state] (2) at (2,0) {$S_{t}=i$};
	\node [state] (3) at (4,0) {$S_{t+1}=i$};
	\node [] (4) at (6,0) {$\ldots$};
	\node [state] (5) at (8,0) {$S_{t+r-1}=i$};
	\node [state] (6) at (10,0) {$S_{t+r}=j$};
    \node [state] (7) at (0,2) {$Y_{t-1}$};
	\node [state] (8) at (2,2) {$Y_{t}$};
	\node [state] (9) at (4,2) {$Y_{t+1}$};
	\node [state] (11) at (8,2) {$Y_{t+r-1}$};
	\node [state] (12) at (10,2) {$Y_{t+r}$};
	\draw[->, line width=0.3pt, black] (1) to (2);
	\draw[->, line width=0.3pt,black] (2) to (3);
	\draw[->, line width=0.3pt,black] (3) to (4);
	\draw[->, line width=0.3pt,black] (4) to (5);
	\draw[->, line width=0.3pt,black] (5) to (6);
    \draw[->, line width=0.3pt,black] (1) to (7);
	\draw[->, line width=0.3pt,black] (2) to (8);
	\draw[->, line width=0.3pt,black] (3) to (9);
	\draw[->, line width=0.3pt,black] (5) to (11);
	\draw[->, line width=0.3pt,black] (6) to (12);
	\draw[decoration={brace,mirror,raise=0.5cm,amplitude=1em},decorate,thick] (2,-0.6) to (8,-0.6);
    \node[] (13) at (5,-2) {\footnotesize{dwell time $r$ drawn from $d_i$}};
    \draw [->] (1) to [out=-60,in=-120] (2);
     \node[] (15) at (1,-1.2) {\footnotesize{$\omega_{ji}$}};   
    \draw [->] (5) to [out=-60,in=-120] (6);
     \node[] (16) at (9,-1.2) {\footnotesize{$\omega_{ij}$}};   
\end{tikzpicture}
\caption{Dependence structure of an HSMM. Whenever the semi-Markov chain enters a new state $i$ at time $t$, the dwell time $r$, i.e.\ the time spent in that state, is drawn from the corresponding dwell-time distribution $d_i(r)$. Consequently, a state switch must occur at time $t+r$ and state $j$ is entered with the conditional probability $\omega_{ij}$. Each observation $Y_t$ depends on the corresponding state $S_t$ and is generated by the associated state-dependent distribution $f_{S_{t}}(y_t)$.}
\label{fig:HSMM}
\end{figure}
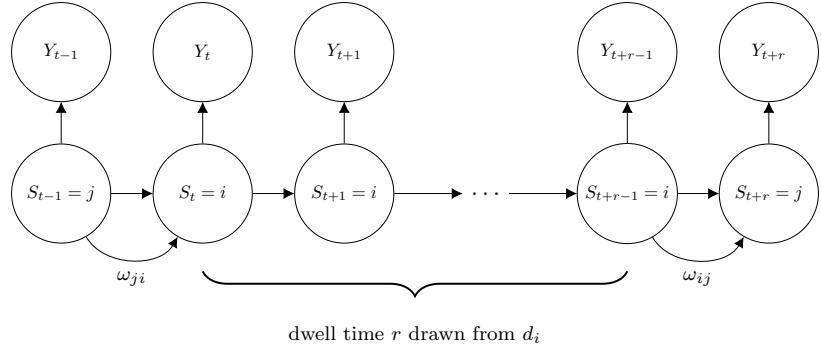
\end{center}

\subsection{Flexible estimation of the state dwell-time distributions}\label{Sec2.2}

\subsubsection{Flexible dwell-time distributions and HMM representation}\label{Sec2.2.1}
Similar to \citet{san01} and \citet{lan11}, we consider dwell-time distributions with an unstructured start and a geometric tail. That is, for each state $i=1,\ldots, N$ and dwell times $r \in \{1,2,\ldots,R_{i}\}$, we assign a parameter $\pi_{i,r}$ to each individual dwell-time probability $d_{i}(r)$, where $R_i$ denotes the upper boundary for the unstructured start. A geometric tail accounts for dwell-times $r>R_{i}$:
$$d_{i}(r)=\begin{cases}
 \pi_{i,r} & \text{if } 0<r\leq R_{i}; \\
  \pi_{i,R_{i}}\left(\ \cfrac{1-\sum_{r=1}^{R_{i}} \pi_{i,r}}{1-\sum_{r=1}^{R_{i}-1} \pi_{i,r}} \right)^{r-R_{i}} & \text{if }r>R_{i},
 \end{cases}
$$
with $0 < \pi_{i,r} < 1$ and $\sum_{r=1}^{R_{i}} \pi_{i,r} < 1$. This allows for a flexible and data-driven shape on the support $\{1,\ldots,R_i\}$ while avoiding a restriction for the dwell-time domain. Usually, only small ranges are considered for the unstructured start (for instance, $R_i=1$ in \citealp{san01}; $R_i \in \{1,2,3\}$ in \citealp{lan11}); for our purposes, however, the upper boundary $R_{i}$ should be chosen large enough to capture the main dwell-time support. This can be explored by initially using large values for $R_{i}$, which can subsequently be replaced by suitable smaller values.

Using a state-space expansion and a suitable block structure in the resulting enlarged TPM, an HSMM with such dwell-time distributions can \textit{exactly} be represented as an HMM \citep{lan11,zuc16}. This opens up the way for the efficient standard HMM machinery for parameter estimation and further inference. In the HMM representation, each HSMM state $i$ is represented by a set of $R_{i}$ sub-states forming a so-called state aggregate $I_i=\{\tilde{i}_{1},\ldots, \tilde{i}_{R_{i}}\}$, which leads to a state space of dimension $\tilde{N}=\sum_{i=1}^{N} R_{i}$. We denote the corresponding HMM Markov chain by $\{\tilde{S_t}\}_{t=1}^{T}$. Each HMM sub-state belonging to the state aggregate $I_i$ is associated with the same state-dependent distribution $f_i(y_t)$ and the corresponding transition probabilities are structured and parameterised such that they exactly mirror the HSMM dwell-time distribution $d_i(r)$. For instance, except for the last sub-state $\tilde{i}_{R_{i}}$ which is associated with the geometric tail, no self-transitions are allowed and the state aggregate can only be traversed through in the indexed order, starting with $\tilde{i}_{1}$. This structure is illustrated in Figure \ref{fig:transition_graph} for a 2-state HSMM. For the HMM transition probabilities within the state aggregates, this implies: $\gamma_{\tilde{i}_{r},\tilde{i}_{r}}=\Pr( \tilde{S}_{t}=\tilde{i}_{r}|\tilde{S}_{t}=\tilde{i}_{r})=0$ and $\gamma_{\tilde{i}_{r},\tilde{i}_{l}}=\Pr(\tilde{S}_{t}=\tilde{i}_{l}|\tilde{S}_{t}=\tilde{i}_{r})=0$ for $r=1,\ldots,R_{i}-1$ and $l \neq r+1$. Furthermore, $\gamma_{\tilde{i}_{R_i},\tilde{i}_{r}}=\Pr( \tilde{S}_{t}=\tilde{i}_{r}|\tilde{S}_{t}=\tilde{i}_{R_i})=0$ for $r \neq R_i$. Thus, most of the transition probabilities are fixed to zero. Further details about the HMM representation are provided in the appendix.

\begin{center}
\begin{figure}[!t]
	\centering
	\begin{tikzpicture}[node distance = 1.5cm]
	\tikzset{state/.style = {circle, draw, minimum size = 55pt, scale = 0.725},every loop/.style={}}
	\node [state] (1) at (-0.55,0) {$\tilde{i}_1$};
	\node [state] (2) at (1.5,0) {$\tilde{i}_2$};
	\node [] (3) at (3,0) {$\ldots$};
	\node [state] (4) at (4.5,0) {$\tilde{i}_{R_{i}}$};
	\node [state] (5) at (1.5,2) {$\tilde{j}_{1}$};
	\node [] (6) at (3,2) {$\ldots$};
	\node [state] (7) at (4.5,2) {$\tilde{j}_{R_{j}}$};
	\draw[->, line width=0.3pt, black] (1) to (2);
	\draw[->, line width=0.3pt,black] (2) to (3);
	\draw[->, line width=0.3pt,black] (3) to (4);
	\draw[->, line width=0.3pt,black] (1) to (5);
	\draw[->, line width=0.3pt,black] (2) to (5);
	\draw[->, line width=0.3pt,black] (4) to (5);
	\draw[->, line width=0.3pt,black] (5) to (6);
	\draw[->, line width=0.3pt,black] (6) to (7);
	\path (4) edge [loop right,<-,line width=0.3pt,black] (4);
    \draw [->] (5) to [out=180,in=90] (1);
	\path (7) edge [loop right,<-,line width=0.3pt,black] (7);
    \draw [->] (7) to [out=120,in=120] (1);
	\end{tikzpicture}
	\caption{Example transition graph illustrating the structure of the HMM-representation for a 2-state HSMM. The actual HSMM states are represented by the state aggregates $I_{i}=\{\tilde{i}_{1}, \ldots, \tilde{i}_{R_{i}}\}$ and $I_{j}=\{\tilde{j}_{1},\ldots,\tilde{j}_{R_{j}}\}$, respectively.}
	\label{fig:transition_graph}
\end{figure}
\end{center}

\subsubsection{Penalised maximum likelihood estimation}\label{Sec2.2.2}

For parameter estimation, we use the HMM representation described above (Section \ref{Sec2.2.1}) and focus on numerical maximisation of the (penalised) log-likelihood. Alternatively, maximum likelihood estimation can be carried out using expectation-maximisation (EM) algorithms specifically tailored for HSMM applications (for example, \citealp{san01,gue03,yu03}). However, they usually assume that a new state is entered at the beginning of the observation period ($t=0$). Besides being unrealistic in some cases, this also impedes stationarity \citep{lan11}. For Bayesian HSMM parameter estimation, see, for example, \citet{eco14}.

Using its HMM representation, the likelihood of the HSMM can efficiently be evaluated using the so-called forward algorithm (see, for example, \citealp{zuc16}). It exploits the fact that the likelihood of an HMM can be written as a matrix product,
$$\mathcal{L}(\boldsymbol{\theta}|y_1,\ldots,y_T)=\bm{\delta}\bm{\Gamma}\mathbf{P}(y_{1})\bm{\Gamma}\mathbf{P}(y_{2})\ldots \bm{\Gamma}\mathbf{P}(y_{T})\bm{1}^\top,$$
where $\bm{\delta}$ is the $\Tilde{N}$-dimensional initial distribution, $\bm{\Gamma}$ is the corresponding $\Tilde{N} \times \Tilde{N}$ TPM (see the appendix for further details on its structure), $\bm{1}$ is an $\Tilde{N}$-dimensional row-vector of ones, and $\mathbf{P}(y_t)$ is an $\tilde{N} \times \tilde{N}$ diagonal matrix containing the state-dependent densities evaluated at $y_t$, 
$$\mathbf{P}(y_t)=\text{diag} \bigl(\underbrace{f_{1}(y_{t}),\ldots,f_{1}(y_{t})}_{R_{1} \text{ times}},\ldots,\underbrace{f_{N}(y_{t}),\ldots,f_{N}(y_{t})}_{R_{N} \text{ times}}\bigr).$$
The forward algorithm corresponds to a recursive calculation of the likelihood with computational costs of order $\mathcal{O}(\Tilde{N}^2T)$, which renders numerical maximisation practically feasible. We denote the corresponding log-likelihood by $\ell(\boldsymbol{\theta}|y_1,\ldots,y_T)=\log(\mathcal{L}(\boldsymbol{\theta}|y_1,\ldots,y_T))$.

To avoid overfitting with respect to the dwell-time PMFs, we enforce smoothness by adding a penalty term for the $m$-th order differences of adjacent state dwell-time probabilities. Thus, for parameter estimation, we maximise the resulting penalised log-likelihood,
$$\hat{\boldsymbol{\theta}} = \underset{\boldsymbol{\theta}}{\text{argmax}} \; \ell(\boldsymbol{\theta}|y_1,\ldots,y_T)-\sum_{i=1}^{N} \lambda_{i} \sum_{r=m+1}^{R_{i}} (\Delta^m \pi_{i,r})^{2},$$
where $\Delta^m \pi_{i,r}$ denotes the $m$-th order difference, $\Delta \pi_{i,r}=\pi_{i,r}-\pi_{i,r-1}$ and $\Delta^{m}=\Delta^{m-1} (\Delta \pi_{i,r})$. There are three types of tuning parameters which influence the estimation. First, the smoothing parameter vector $\bm{\lambda}=(\lambda_{1},\ldots,\lambda_N)$ controls the balance between goodness-of-fit and smoothness of the dwell-time PMFs $d_{i}(r)$. For $\bm{\lambda}=\bm{0}$, the penalty term completely disappears from the equation and the estimation reduces to a simple maximum likelihood estimation. Since in general, the different states' dwell-time distributions require different degrees of smoothing, the smoothing parameters are chosen for each state individually, i.e.\ $\lambda_i \neq \lambda_j$ for $i \neq j$ is possible. A common way to select the smoothing parameters is via cross validation (see \citealp{lan15, ada19}). Second, the difference order $m$ influences the shape of $d_{i}(r)$, especially when $\lambda_{i}$ becomes large. For instance, for $m=1$ and $\lambda_{i} \rightarrow \infty$, $d_i(r)$ approaches a uniform distribution, while for $m=2$ and $\lambda_{i} \rightarrow \infty$, $d_i(r)$ approaches a distribution with a linearly decreasing PMF. Higher-order differences can result in more flexible distributional shapes. We recommend a pragmatic choice of $m$ based on the data at hand, the results arising from an initial unpenalised estimation and a close inspection of the goodness of fit. Similar to \citet{ada19}, we made the experience that $m \ge 3$ provides a reasonable choice in many applications. Third, the upper boundary $R_{i}$ determines the range for which  $d_{i}(r)$ is explored. If chosen too small, the estimation might miss important patterns of the dwell-time distribution. If chosen very large, numerical instabilities might arise (especially for small $\lambda_{i}$), the required memory increases and the computational costs become demanding. A simple and pragmatic approach to find suitable boundary values for the unstructured start is to carry out an initial estimation with large values for $R_{i}$, $i=1,\ldots,N$, and no penalisation, i.e.\ $\bm{\lambda}=\bm{0}$. This provides first insights about the core dwell-time support which can then be used to adjust $R_i$ accordingly.

\section{Case study: Investigating dwell times in muskox movements}\label{Sec3}
\label{Sec3}

We illustrate our PML approach using real GPS-based muskox (\textit{Ovibos moschatus}) movement data. For HMMs, movement ecology is an important area of application with the states usually being interpreted as proxies for the animals' unobserved behavioural modes driving the observed movement patters \citep{mcc20}. Similarly, HSMMs with parametric (e.g.\ shifted Poisson and negative binomial) dwell-time distributions have successfully been applied in this context \citep{lan12,lan14,van15}. For muskox movements in northeast Greenland, \citet{beu20} found that a 3-state HMM adequately describes the main behavioural states `resting', `foraging', and `relocating'. They applied the model to step length (metre) and turning angle (radian) based on hourly GPS locations. While \citet{beu20} account for temporal variation in the transition probabilities using environmental covariates, here, we focus on the direct estimation of the state dwell-time distribution. As ruminants, muskoxen need to forage and rest on a regular basis. Thus, the explicit estimation of the states' dwell-time distributions could provide new insights into the animals' behavioural patterns, in particular into the durations of foraging and resting bouts.

For simplicity, we consider the movement track from a single muskox during the winter season 2013/14 with length $T=6825$ (including $6769$ registered GPS locations and $56$ missing locations), a subset of the data used by \citet{beu20}.
\begin{figure}[h!t]
    \centering
    \includegraphics[width=0.8\textwidth]{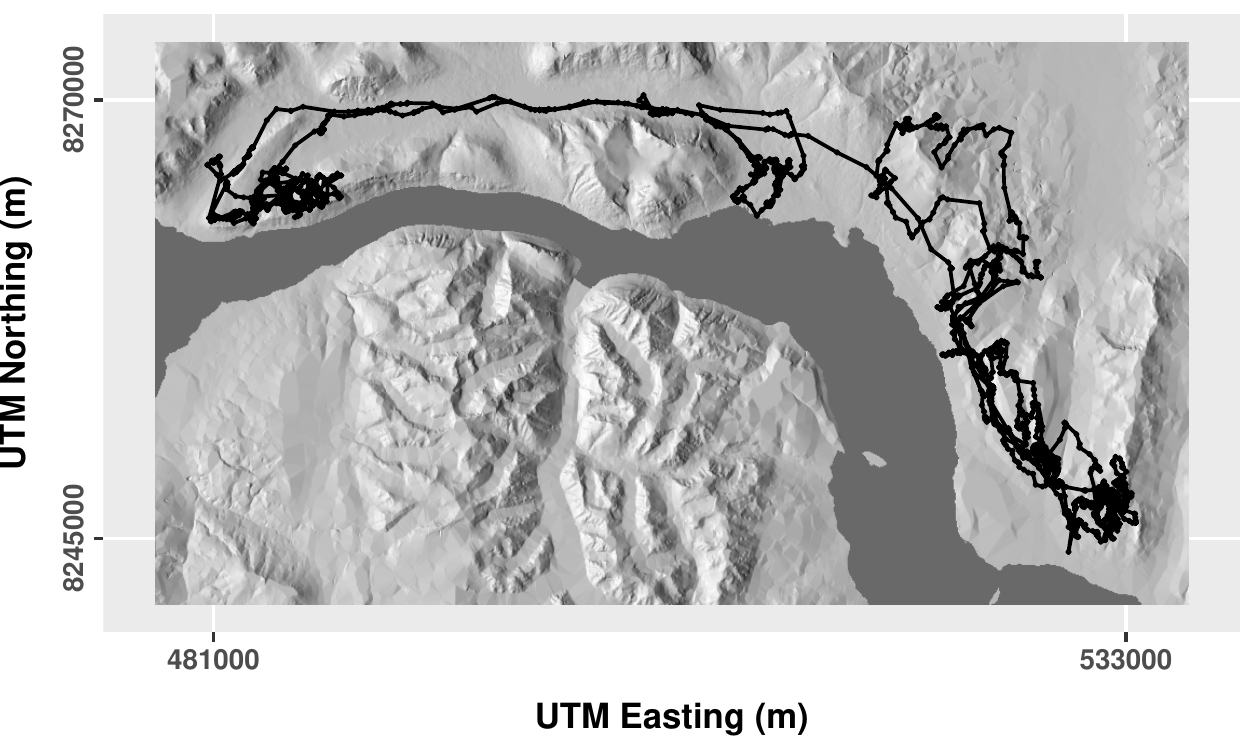}
    \caption{Recorded muskox movement track based on hourly GPS locations.}
    \label{fig:track}
\end{figure}
The movement track is displayed in Figure  \ref{fig:track}. Assuming contemporaneous conditional independence, we consider a 3-state HMM and 3-state PML-based HSMMs, hereafter denoted as PML-HSMMs, with state-dependent gamma distributions for step length and von Mises distributions for turning angle. This is in line with the analysis of \citet{beu20}. To account for the zero step length observations included in the data, we consider additional parameters corresponding to point masses on zero. The tuning parameters $R_{i}$ within the PML-HSMM are selected based on a preliminary unpenalised estimation ($\bm{\lambda}=\bm{0}$) using $30$ freely estimated dwell-time probabilities for each state, respectively (i.e.\ $R_1=R_2=R_3=R=30$). The resulting PMFs are displayed in Figure S1 in the Supplementary Material, indicating that dwell times $r \le 10$ capture most of the probability mass for all three states ($98.24 \%$, $98.74 \%$, and $94.73 \%$ for state 1, 2, and 3, respectively). This is also biologically reasonable as the muskox is generally expected to switch its behavioural modes during the day. Thus, for our analysis, we use an unstructured start of length $R=10$ for all states. To ensure enough flexibility for the dwell-time distributions, we penalise the $4$-th order differences ($m=4$). However, in the Supplementary Material, we provide results arising from $m \in \{1,2,3\}$ using $R=10$, and $R \in \{5,20\}$ using $m=4$, to provide information about the sensitivity of these choices. All models were fitted in \texttt{R} \citep{rco20} using the numerical optimisation procedure \texttt{nlm}. To speed up estimation, the forward algorithm was implemented in \texttt{C++}.

\begin{figure}[!t]
    \centering
    \includegraphics[width=0.95\textwidth]{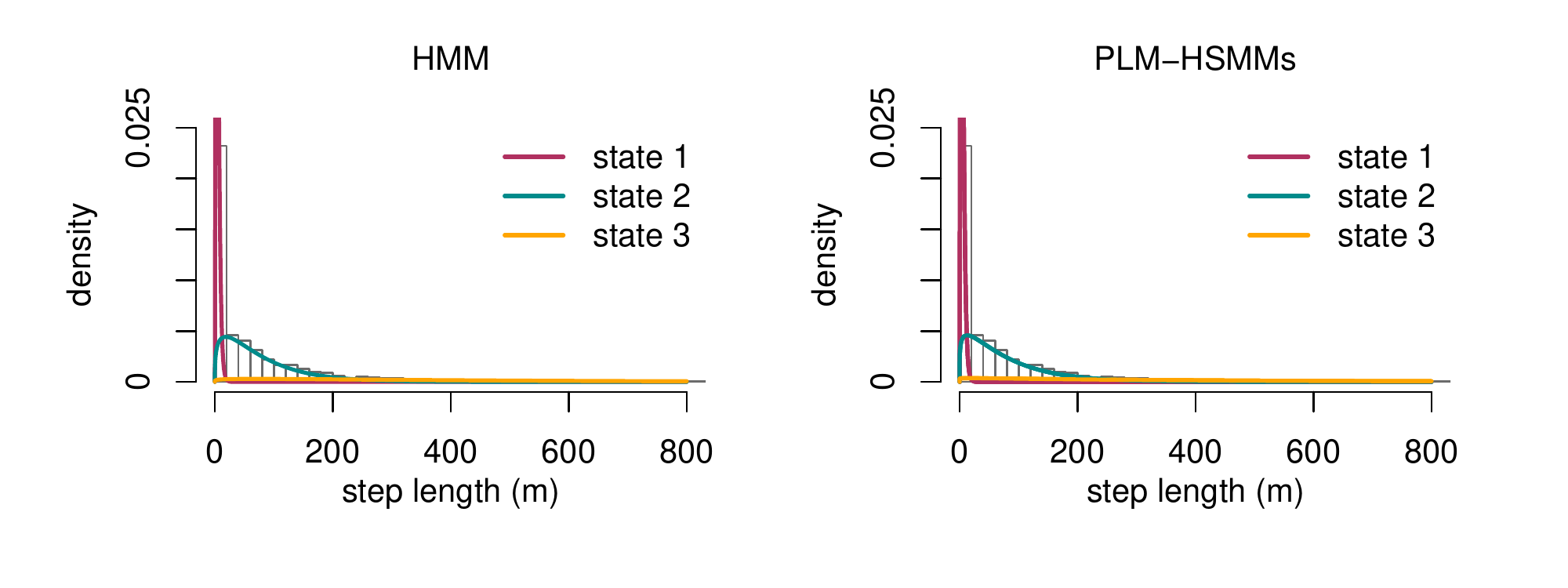}
    
    \vspace{-3em}
    
    \includegraphics[width=0.95\textwidth]{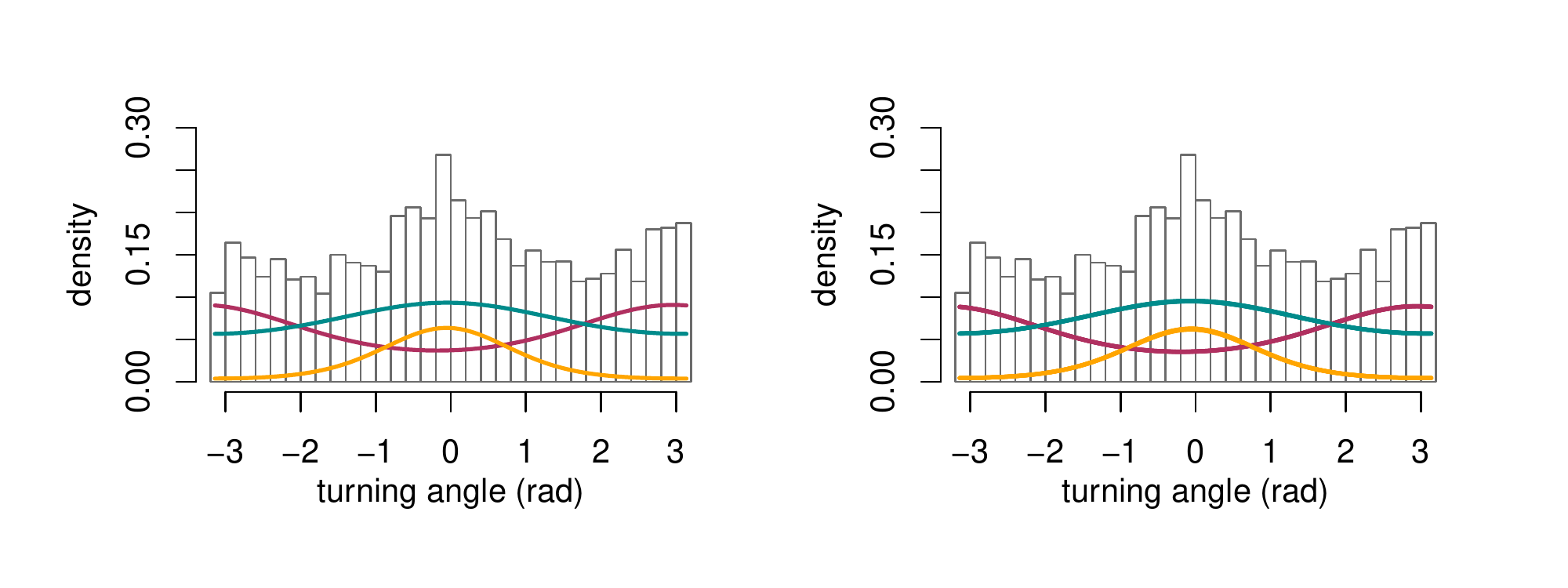}
    \caption{Estimated state-dependent gamma distributions for step length and von Mises distributions for turning angles, resulting from the 3-state models considered. The left panels show the results of the HMM. The right panel shows the results of all PML-HSMMs for which the distributions resulting from different choices of $\lambda$ are plotted on top of each other. It is, however, difficult to see any differences between the results of the different PML-HSMMs, because the corresponding estimates are very similar to each other. All distributions are weighted by the stationary distribution and the background shows the corresponding histograms of the observed variables.}
    \label{fig:sdd}
\end{figure}

\begin{figure}[!t]
    \centering
    \includegraphics[width=\textwidth]{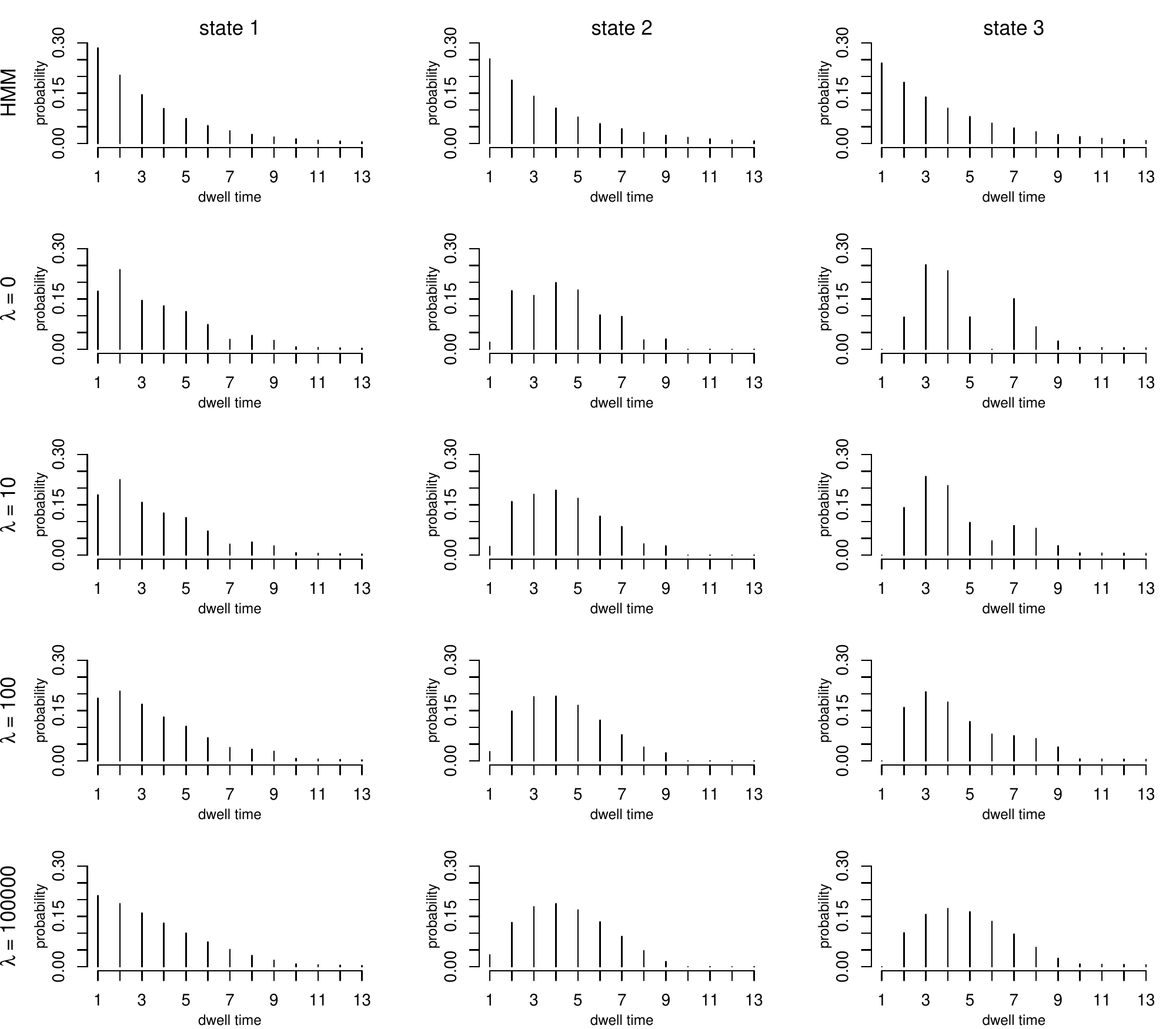}
    \caption{Estimated dwell-time distributions of the 3-state HMM and 3-state PML-HSMMs using different smoothing parameter values $\lambda$.}
    \label{fig:dwell_time_sl}
\end{figure}

To demonstrate the effect of the penalisation, we first present results from simplified PML-HSMMs with $\lambda_{1}=\lambda_{2}=\lambda_{3}=\lambda$ and $\lambda \in \{0,10^1,10^2,10^5\}$. Figure \ref{fig:sdd} shows the estimated state-dependent gamma distributions (for step length) and von Mises distributions (for turning angle) resulting from the fitted 3-state HMM and PML-HSMMs, respectively. The state-specific patterns are very similar across the models and comparable to the results of \citet{beu20}. Thus, the states can reasonably be interpreted as corresponding roughly to resting (state 1), foraging (state 2), and relocating (state 3), respectively.
The dwell-time distributions, however, are very different across the fitted models, as displayed in Figure \ref{fig:dwell_time_sl}. Regardless of the choice of $\lambda$, the estimated PML-HSMM dwell-time distributions differ substantially from geometric distributions, especially for state 2 and 3 where the modal dwell time is clearly greater than one. This suggests that a basic HMM would not correctly represent the dynamics in the state process. The necessity of penalisation becomes clear for example in view of $\hat{d}_3(r)$, the dwell-time distribution estimated for state 3: when increasing $\lambda$, the distribution becomes smoother, and in particular the gaps in the PMF, as obtained when not penalising ($\lambda=0$; top right panel in Figure \ref{fig:dwell_time_sl}), are filled due to the enforced smoothness. With a strong penalisation using $\lambda=10^5$, even the second mode in $\hat{d}_3(r)$ diminishes (bottom right panel), which otherwise appears when using the smaller smoothing parameter values. Note that especially for large values of $\lambda$, the shape of the smoothed PMFs depends on the choice of the difference order $m$. This is illustrated in the Supplementary Material where Figures S2--S4 display the dwell-time distributions resulting from $m=1,2,3$, respectively. While for $\lambda=10^1$ and $\lambda=10^2$, the results are comparable across the choice of $m$, for $\lambda=10^5$, the estimated dwell-time distributions greatly differ. For instance, the PMFs approach uniform distributions on $r \le 10$ when penalising the first-order differences ($m=1$, Figure S2) and linearly decreasing distributions using the second-order differences ($m=2$, Figure S3). Based on the biological context and the results from $\lambda=0$, both do not seem to be appropriate in this case study. We expect this to be the case for most applications.

To find an appropriate model for the muskox movement data, we carried out a two-step model selection procedure: (i) To select an appropriate vector $\bm{\lambda}=(\lambda_{1},\lambda_{2},\lambda_{3})$ for the PML-HSMM, we used a $10$-fold cross validation based on the neighbourhood algorithm proposed by \citet{lan15} with scores being the averaged log-likelihood across the validation samples. With the focus being on the dwell-time distributions, we used a blockwise partitioning of the data and considered a $3$-dimensional grid of powers of tens, i.e.\ $\{10^0,10^1,10^2,\ldots\}^3$. This resulted in the selection of $\bm{\lambda}=(10^5,10^4,10^2)$. (ii) The HMM, HSMM with negative binomial distribution, and PML-HSMM with $\bm{\lambda}=\bm{0}$ form a set of natural candidate models for the PML-HSMM selected via cross validation. We used AIC to select among these candidate models, where for the PML-HSMM, we approximated the effective degrees of freedom using the trace of the empirical Fisher matrix of the unpenalised model ($\bm{\lambda}=\bm{0}$) multiplied by the Fisher matrix of the penalised model with $\bm{\lambda}=(10^5,10^4,10^2)$ (following the approach of \citealp{gra92}; see also \citealp{lan18}).
For estimation, the 3-state HSMM with negative binomial distribution was approximated by an HMM as proposed by \citet{lan11} with state aggregates of dimension $30$ per HSMM state. The resulting AIC values are displayed in Table \ref{tab:AIC}. The PML-HSMM is clearly preferred over both the HMM and the negative binomial HSMM. According to the AIC, the best model among the candidate models is the PML-HSMM with $\bm{\lambda}=(10^5,10^4,10^2)$.

\begin{figure}[t!]
    \centering
    \includegraphics[width=\textwidth]{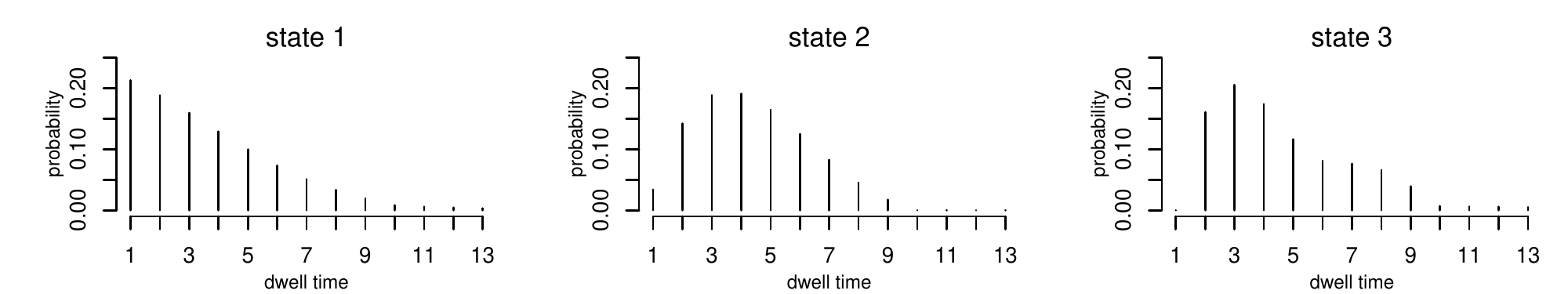}
    \caption{Estimated dwell-time distributions of the 3-state PML-HSMM selected by cross validation with smoothing parameter vector $\bm{\lambda}=(10^5,10^4,10^2)$.}
    \label{fig:dwell_time_cv}
\end{figure}

The corresponding dwell-time distributions are displayed in Figure \ref{fig:dwell_time_cv}. The results suggest that the tracked muskox tends to forage and travel for several hours before switching to a different state, with modal values being $r=4$ and $r=3$, respectively. However, $\hat{d}_3(r)$ seems to be almost bimodal, indicating that there might be different types of travelling periods, i.e.\ long and short travelling phases. This distributional shape would not have been captured by standard parametric HSMMs. The modal dwell time for state 1 (resting) is $r=1$, but with a rather slow decay compared to the geometric distribution. Thus, the resting periods tend to be slightly shorter than the foraging and relocation periods and tend to last only a few hours. A pseudo-residual analysis is provided in Section 2 of the Supplementary Material, indicating a good model fit for the selected PML-HSMM.

\begin{table}[t!]
    \centering
    \begin{tabular}{l|cccc}
    \toprule
    model                       & no.\ par.\ / df   & $\ell$     & AIC    & $\Delta$ AIC \\\midrule
    HMM                         &       21          &      -44964.04    &    89970.07  &  231.31 \\ 
    nbHSMM                      &       24          &      -44897.09    &    89842.18  &  103.41 \\
    PML-HSMM$_{(0,0,0)}$            &       48          &      -44823.71    &    89743.43  &  4.66\\ 
    PML-HSMM$_{(10^5,10^4,10^2)}$ &       32.70       &      -44835.96    & \textbf{89737.32}  &  0\\ 
         \bottomrule
    \end{tabular}
    \caption{Number of parameters/effective degrees of freedom, log-likelihood values, AIC values and $\Delta$ AIC for  the 3-state models considered.}
    \label{tab:AIC}
\end{table} 
%

\section{Discussion and conclusions}\label{Sec4}

As the state process is unobserved, it is often unclear how to select a model that appropriately reflects the underlying state dynamics. We introduced a penalised estimation approach which combines PMFs with an unstructured start and higher-order difference penalties to derive flexible yet smooth estimates for the states' dwell-time distributions. While HSMMs with standard parametric distributions are in general more parsimonious than PML-HSMMs, they are restricted in their distributional shapes and therefore might fail in capturing the underling dwell-time patterns. For instance, consider the negative binomial distribution shifted by one, which comprises the geometric distribution as a special case (with shape parameter equal to one). Thus, to some extent, negative binomial HSMMs actually allow for different shapes, can identify states for which geometric dwell-time distributions suffice \citep{gue05} and can be tested against the nested HMMs \citep{bul06}. However, they are not able to identify more complex patterns like bimodal dwell-time distributions. Avoiding strong distributional assumptions, our penalised estimation approach can be used as an exploratory tool to investigate the unknown shapes of the states' dwell-time distributions. The method can either serve for direct modelling purposes, or as a basis for subsequent modelling choices, for example, in order to decide whether an HMM would be appropriate for the data at hand, or what distributional assumption may be adequate within a conventional HSMM (in the spirit of \citealp{san01}). Thereby, it could also indicate if different states require different families of parametric distributions.

Due to the HMM representation, inference is straightforward and can completely rely on well-known HMM techniques \citep{lan11}. This is in line with \citet{joh05} who, based on a comparison of different algorithms and HSMM-like model formulation, argues that the use of standard models with special state topologies is practically more reasonable than the use of more complex and expensive algorithms. The HMM representation makes it fairly easy to change the distributional assumption in the state-dependent process and to adapt the model to the application at hand. Only when the number of states or the number of sub-states in the state aggregates becomes large, the likelihood evaluation might suffer from the use of large matrices and the memory required. An alternative approach would be the implementation of an EM algorithm with a roughness penalty term which is shortly discussed by \citet{gue03} for HSMMs with non-parametric dwell-time distributions.

The PML-HSMM approach allows for a straightforward incorporation of covariates into the state-dependent process \citep{lan11}. However, as for HSMMs in general, it is conceptually unclear how to integrate covariates into the state process of the model. Especially in movement ecology, the interest often lies in the influence of environmental variables on the animal's movement behaviours (for example, \citealp{vbe19,beu20,pho20}). Within HMMs, the transition probabilities and covariates can be linked via (multinomial) logit link functions \citep{zuc16}. Thus, depending on the covariate values, the transition probabilities change over time. This also affects the probability to remain in the current state and consequently, the implicit states' dwell-time distributions. While in principle, the conditional transition probabilities of an HSMM can be linked to covariates in the same way, this would not directly affect the dwell-time distributions of the model as within an HSMM, the dwell-time distributions are modelled separately from the conditional transition probabilities. Alternatively, the HSMM parameters defining the dwell-time distributions could be linked to covariates. But as the time at which the state process enters a new state is unknown, it is unclear on which covariate observations the dwell-time parameters should depend on. Therefore, if the interest of the analysis lies on the influence of time-varying covariates on the state process, HMMs provide a more convenient framework. However, in cases where covariates are not assumed to influence the state process, or where no covariates are available, the proposed PML-HSMM approach can provide new insights into the states' dwell-time distributions and the underlying latent state dynamics. For univariate time series and common state-dependent distributions, the PML-HSMM approach is implemented in the \texttt{R} package \texttt{PHSMM} \citep{poh21} on CRAN.

\section*{Acknowledgements}

The authors are very grateful to Roland Langrock for inspiring and valuable discussions and helpful advice that considerably improved the paper. They also thank Niels Martin Schmidt for providing the muskox tracking data.

\renewcommand\refname{References}
\makeatletter
\renewcommand\@biblabel[1]{}

\section*{Appendix}\label{app}

Here we describe the structure of the HMM which exactly represents the $N$-state HSMM described in Section \ref{Sec2.2.1}, following the approach of \citet{lan11}. For each state $i = 1,\ldots,N$, the dwell-time distribution $d_{i}(r)$ of the considered HSMM is defined by an unstructured start for duration $r \in \{1,\ldots,R_i\}$, $R_i \in \mathbb{N}$, and a geometric tail (see Section \ref{Sec2.2.1} for details). The conditional transition probabilities are summarised in the matrix $\Omega=(\omega_{ij})$ with $\omega_{ij}=\Pr(S_{t}=j|S_{t-1}=i,S_{t} \neq i)$ for $i \neq j$ and $\omega_{ii}=0$, adhering the row-constraints $\sum_{i=1}^N \omega_{ij}=1$. Expanding the state space, the HSMM can exactly be represented by an HMM with state space of dimension $\tilde{N}=\sum_{i=1}^{N} R_{i}$ and we denote the corresponding Markov chain sub-states by $\tilde{S}_t$. The HMM sub-states are organised in the so-called state aggregates $I_i=\{\tilde{i}_{1},\tilde{i}_{2},\ldots,\tilde{i}_{R_i}\}$, $i=1,\ldots,N$, where state aggregate $I_{i}$ represents the HSMM state $i$. Consequently, all sub-states belonging to state aggregate $I_i$ are associated to the same state-dependent distribution: $f(y_{t}|\tilde{S}_t \in I_i)=f(y_t|S_t=i)$.
To reproduce the HSMM dwell-time distributions and state-switching patterns, the $\tilde{N} \times \tilde{N}$ TPM $\Gamma$ of the HMM is organised in a block-structure:
$$\Gamma=\begin{pmatrix}
    \Gamma_{11} & \ldots & \Gamma_{1N} \\
    \vdots & \ddots & \vdots \\
    \Gamma_{N1} & \ldots & \Gamma_{NN}
    \end{pmatrix}$$
The diagonal block elements $\Gamma_{ii}$, $i=1,\ldots,N$, are of dimension $R_{i} \times R_{i}$ and represent the dwell-time distributions $d_{i}(r)$. For $R_i \ge 2$, they are structured as follows:
$$\Gamma_{ii}=\begin{pmatrix}
    0       & 1-c_i(1)  & 0         & \ldots    & 0     \\
    0       & 0         & \ddots    & \ddots    & \vdots \\
    \vdots  & \vdots    &           &           & 0     \\
    0       & 0         & \ldots    & 0         & 1-c_i(R_i-1) \\
    0       & 0         & \ldots    & 0         & 1-c_i(R_i) \\
    \end{pmatrix},$$
with $c_i(r)=\cfrac{d_i(r)}{1-F_i(r-1)}$ for $r=1,\ldots,R_i$, and $F_i$ denotes the cumulative distribution function $F_i(r)=\sum_{k=1}^{r} d_i(k)$ associated to state $i$. In case of $R_i=1$, $\Gamma_{ii}= 1-c_i(1)$ and the state dwell-time distribution becomes a geometric distribution. This case, however, is not explicitly considered in this paper.

The $R_i \times R_j$ off-diagonal block elements $\Gamma_{ij}$, $i \neq j$, represent the state-switching probabilities and are structured as follows:
$$\Gamma_{ij}=\begin{pmatrix}
    \omega_{ij}c_i(1) & 0   &   \ldots & 0 \\
    \omega_{ij}c_i(2) & 0   &   \ldots & 0 \\
    \vdots          &   \vdots    & \ddots   & \vdots \\
    \omega_{ij}c_i(R_i) & 0 & \ldots & 0 \\
    \end{pmatrix}.$$
In case of $R_j=1$, the columns of zeros disappear, but again, this case is not explicitly considered in this paper.

\end{spacing}

\end{document}


\begin{spacing}{1.5}
\maketitle

\renewcommand{\thefigure}{S\arabic{figure}}

\section{Additional PML-HSMM results for the muskox case study}

\subsection{Main dwell-time support}
\begin{figure}[h]
    \centering
    \includegraphics[width=\textwidth]{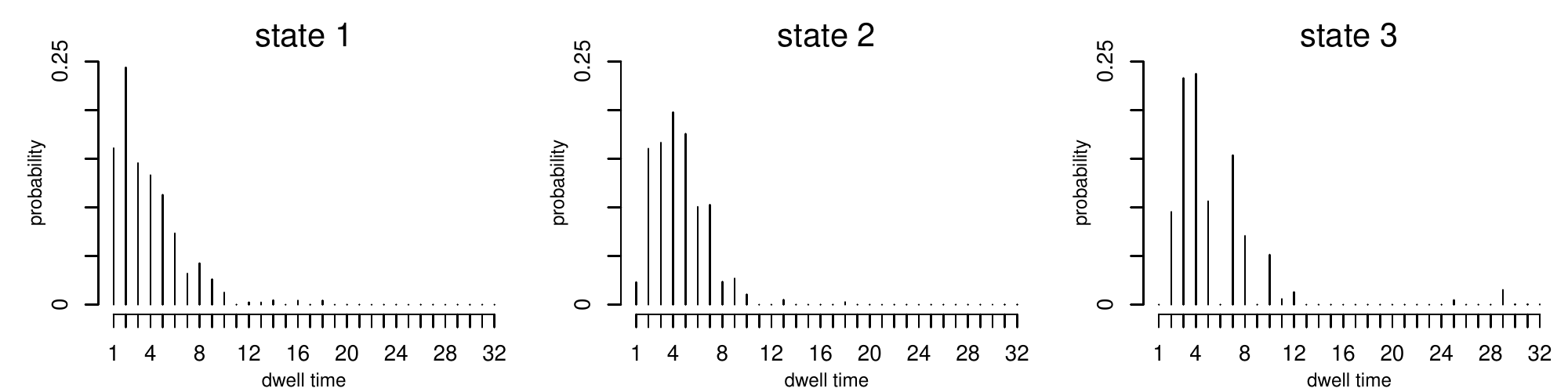}
    \caption{Estimated dwell-time distributions of the 3-state PLM-HSMM with $\lambda=0$ and an unstructured start of length $R=30$ for each state.}
    \label{fig:d_i_l0}
\end{figure}

In this section, we provide supplementary results for the muskox movement data discussed in Section 3 in the main manuscript. Figure \ref{fig:d_i_l0} displays the estimated 3-state PML-HSMM dwell-time distributions using $\lambda_{1}=\lambda_{2}=\lambda_{3}=\lambda=0$ and $R_{1}=R_{2}=R_{3}=R=30$. It indicates that the core dwell-time support for the muskox movement data is $\{1,\ldots,10\}$. For $r > 10$, most dwell-time probabilities are estimated very close to zero. This result is the basis for setting $R=10$ for the subsequent analysis.

\vspace{5em}

\newpage
\subsection{Influence of the difference order}
\begin{figure}[hb]
    \centering
    \includegraphics[width=\textwidth]{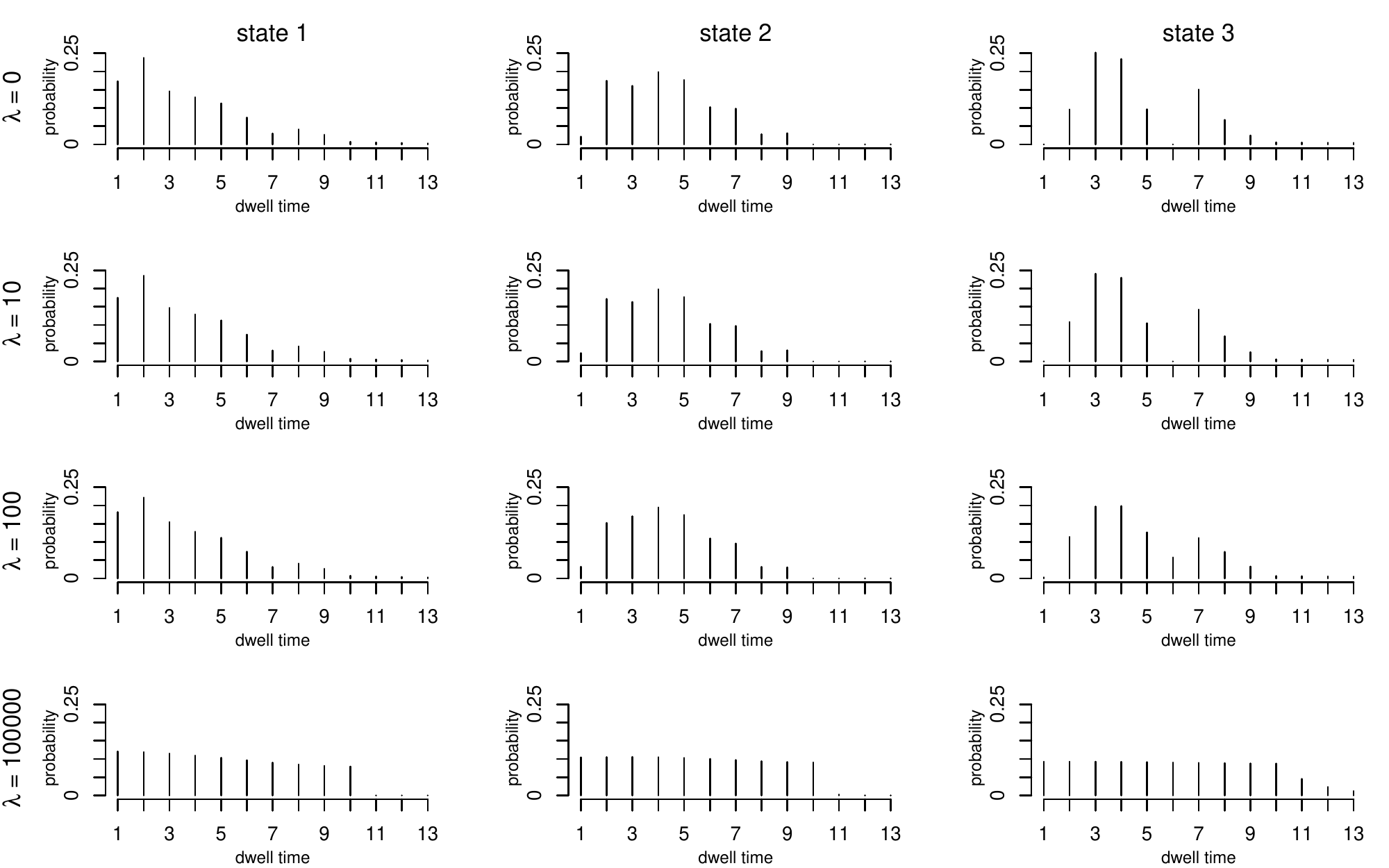}
    \caption{Estimated dwell-time distributions of the 3-state PLM-HSMMs with penalisation of first-order differences ($m=1$) and an unstructured start of length $R=10$ for each state.}
    \label{fig:d_i_df1}
\end{figure}

\begin{figure}[t]
    \centering
    \includegraphics[width=\textwidth]{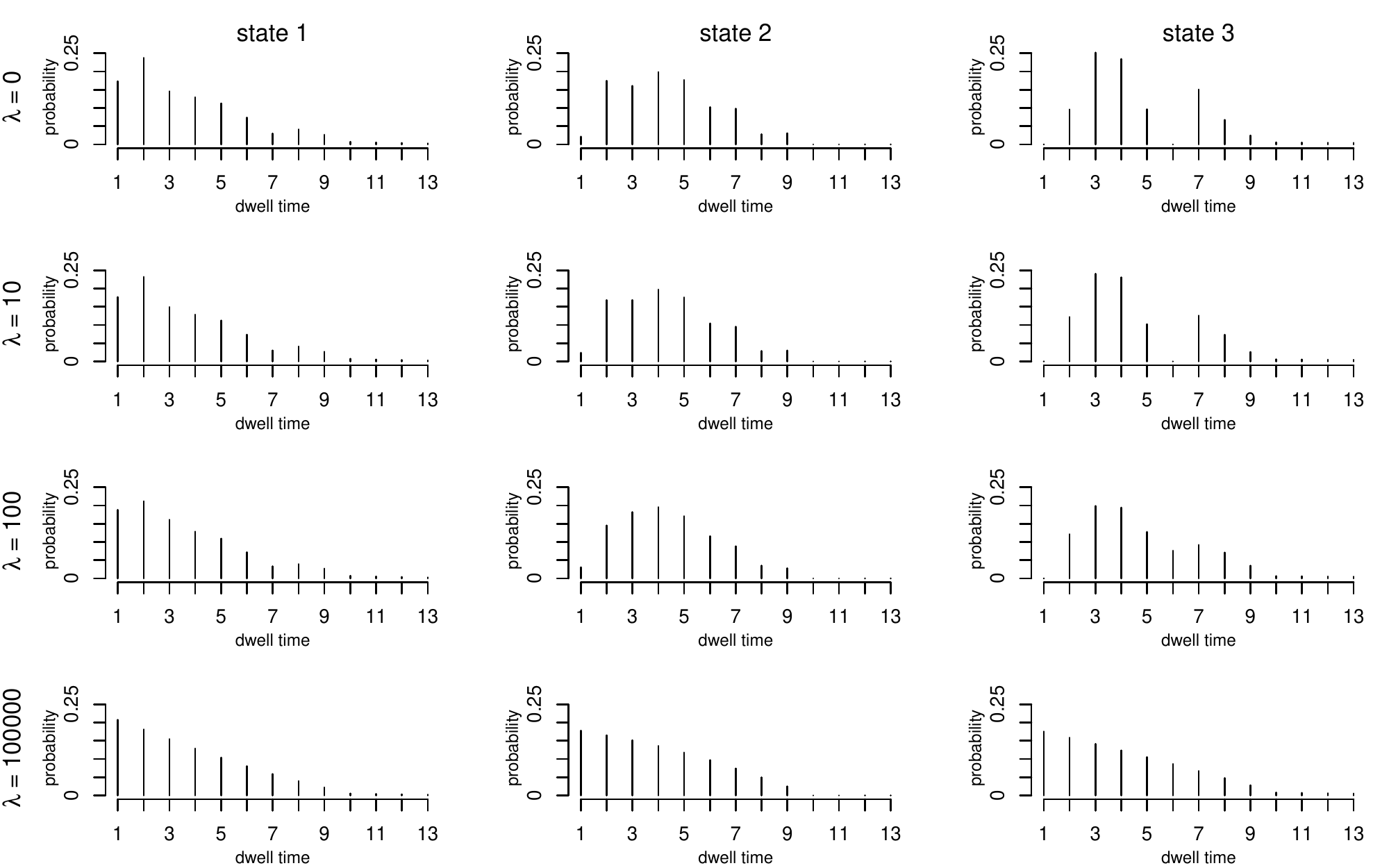}
    \caption{Estimated dwell-time distributions of the 3-state PLM-HSMMs with penalisation of second-order differences ($m=2$) and an unstructured start of length $R=10$ for each state.}
    \label{fig:d_i_df2}
\end{figure}

\begin{figure}[t]
    \centering
    \includegraphics[width=\textwidth]{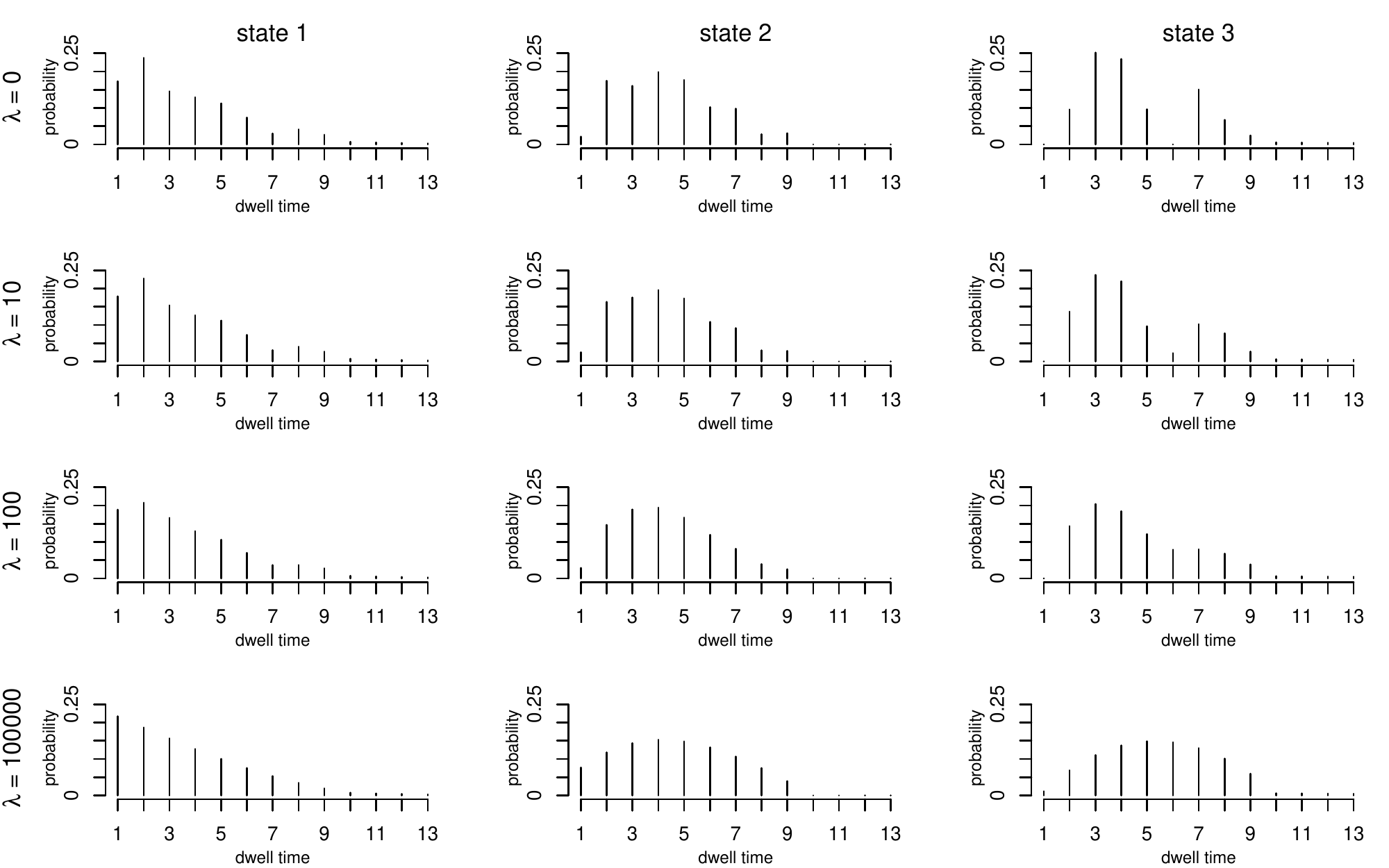}
    \caption{Estimated dwell-time distributions of the 3-state PLM-HSMMs with penalisation of third-order differences ($m=3$) and an unstructured start of length $R=10$ for each state.}
    \label{fig:d_i_df3}
\end{figure}

To illustrate the influence of the choice of the difference order $m$ on the resulting dwell-time distributions, for $\lambda \in \{10^1,10^2,10^5\}$ and $R=10$, we additionally fitted 3-state PML-HSMMs with $m=1,2,3$ to the muskox movement data. The resulting probability mass functions (PMFs) are displayed in Figures \ref{fig:d_i_df1}, \ref{fig:d_i_df2}, and \ref{fig:d_i_df3}, respectively. For $\lambda=10^1$ and $\lambda=10^2$, the fitted dwell-time distributions only differ slightly across the choice of $m$. However, using a strong difference penalisation, i.e.\ $\lambda=10^5$, the difference order $m$ clearly affects the shape of the estimated distributions: For $m=1$, the PMFs approach uniform distributions on $r \le 10$ (with a geometric tail for $r >10$; Figure \ref{fig:d_i_df1}, bottom panel), for $m=2$, a linearly decreasing distribution is approached (with a geometric tail for $r >10$; Figure \ref{fig:d_i_df2}, bottom panel). When penalising third-order differences ($m=3$), the estimated PML-HSMM is able to capture more complex patters.  For instance, the estimated distributions $\hat{d}_1(r)$ and $\hat{d}_2(r)$ differ in their shapes (Figure \ref{fig:d_i_df3}, bottom panel).

\vspace{5em}

\clearpage
\subsection{Length of the unstructured start}
\begin{figure}[hb]
    \centering
    \includegraphics[width=\textwidth]{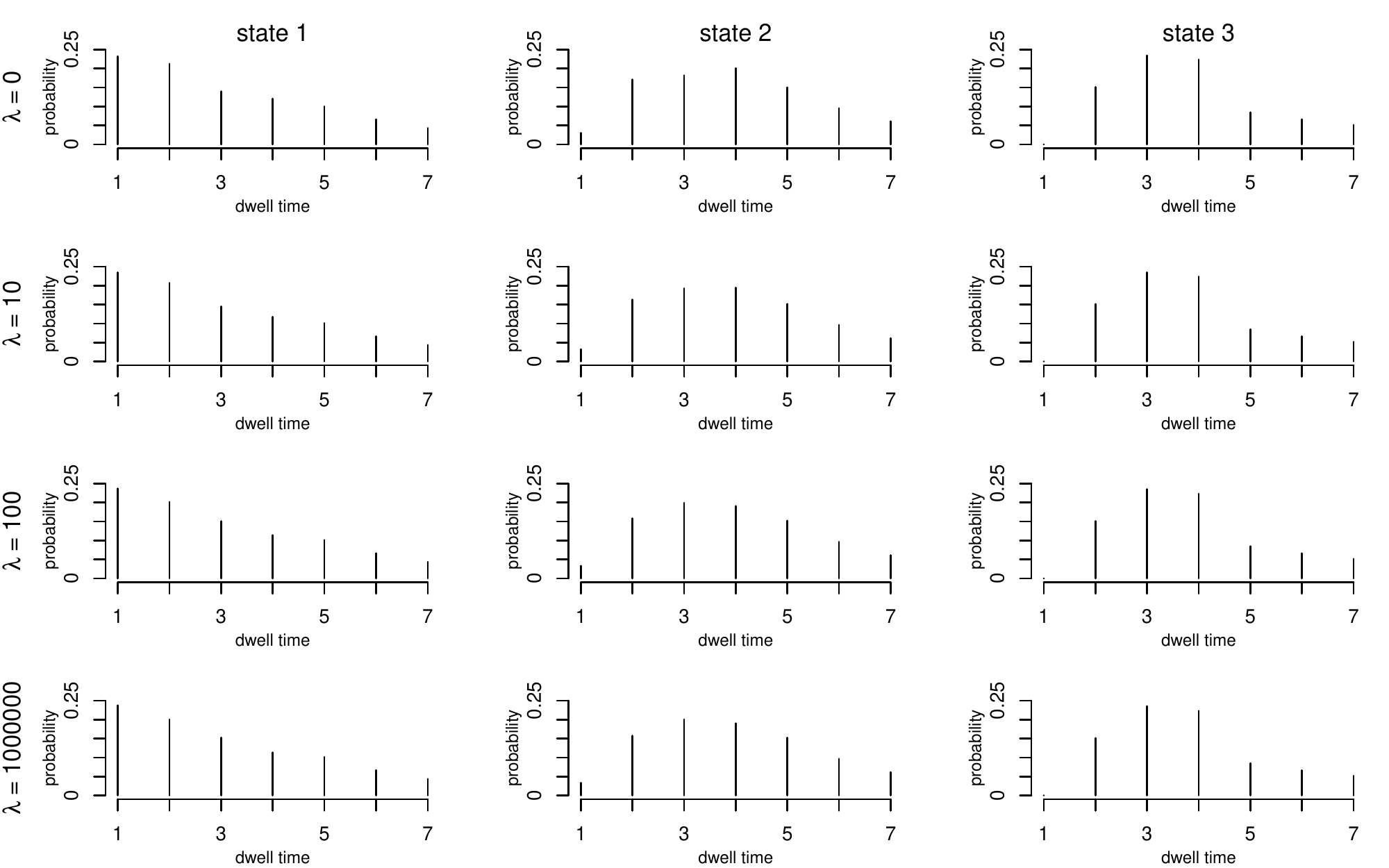}
    \caption{Estimated dwell-time distributions of the 3-state PLM-HSMMs with penalisation of fourth-order differences ($m=4$) and an unstructured start of length $R=5$ for each state.}
    \label{fig:d_i_5df4}
\end{figure}

\begin{figure}[H]
    \centering
    \includegraphics[width=\textwidth]{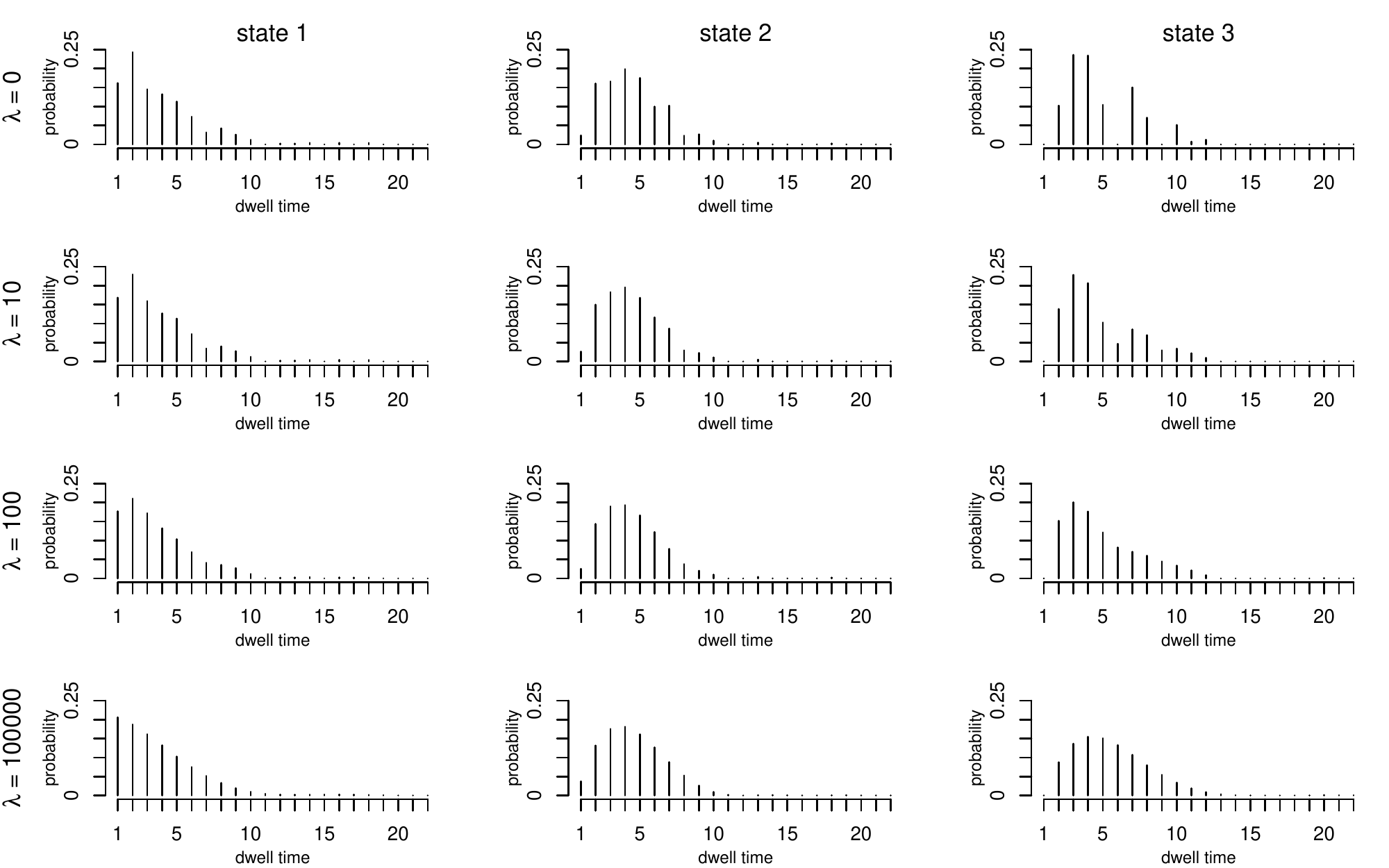}
    \caption{Estimated dwell-time distributions of the 3-state PLM-HSMMs with penalisation of fourth-order differences ($m=4$) and an unstructured start of length $R=20$ for each state.}
    \label{fig:d_i_20df4}
\end{figure}

To illustrate the influence of the length of the unstructured start, for $\lambda \in \{10^1,10^2,10^5\}$ and $m=4$, we further estimated 3-state PML-HSMMs using $R=5$ and $R=20$, respectively. The resulting PMFs are displayed in Figures \ref{fig:d_i_5df4} and \ref{fig:d_i_20df4}. With $R=5$, the penalisation has hardly any effect on the estimation. This can partly be explained by the fact that the unpenalised estimation ($\lambda=0$; Figure \ref{fig:d_i_5df4}, top panels) already results in smooth PMFs and the geometric tails carry a considerable share of the probability masses, namely $19.33\%$, $26.46\%$, and $30.50\%$ in state 1, 2, and 3, respectively. Furthermore, for only $R=5$ freely estimated probabilities, the difference order $m=4$ is chosen too large.

The PMFs resulting from $R=20$ are comparable to the ones arising from setting $R=10$ (Figure 5 in the main manuscript). Hence, the choice of $R=10$ seems suitable for the muskox case study, as it captures the main patterns, is more parsimonious than $R=20$ and computationally less expensive.

\vspace{5em}

\clearpage
\section{Step length pseudo-residuals for the selected PML-HSMM}

\begin{figure}[H]
    \centering
    \includegraphics[width=\textwidth]{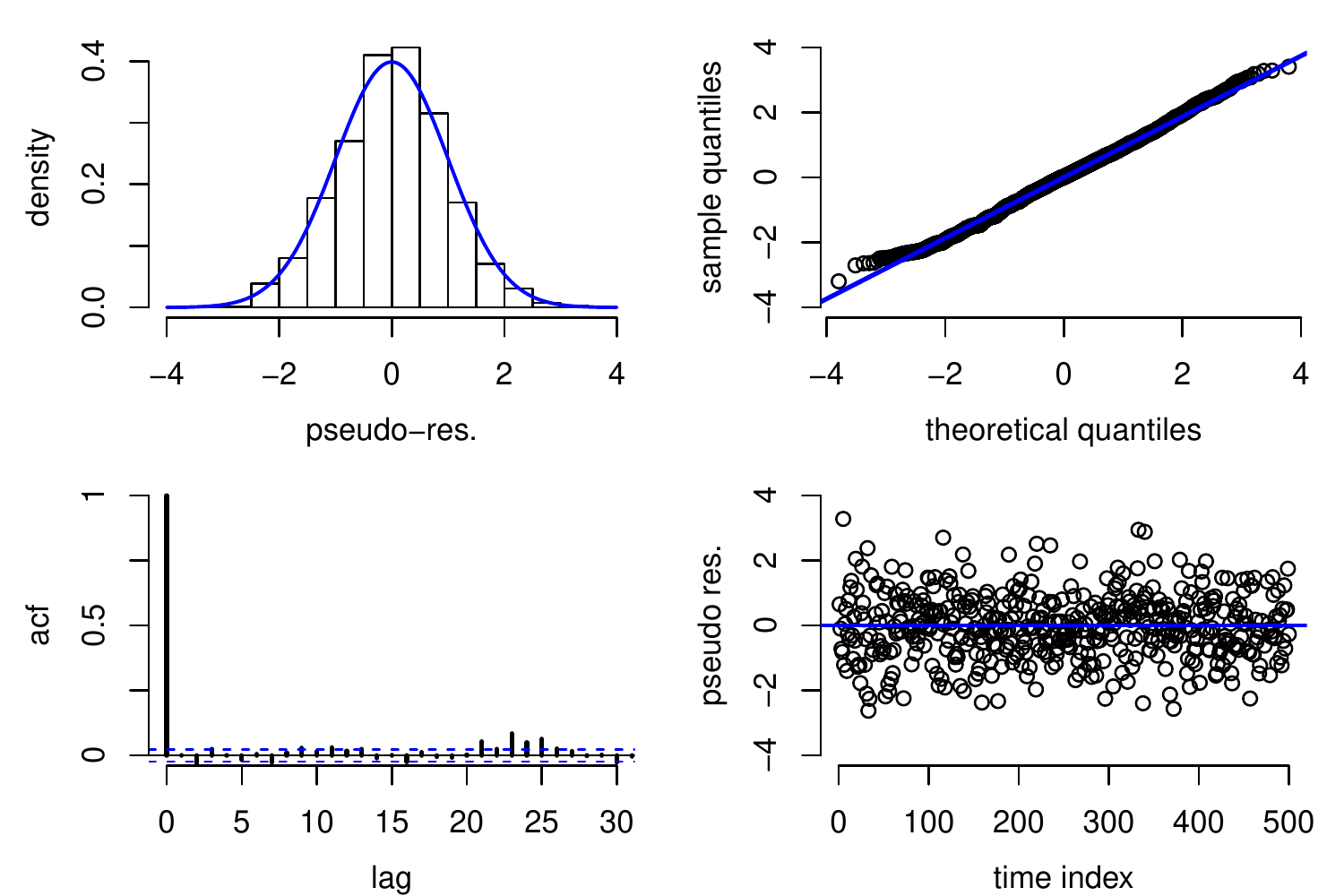}
    \caption{Ordinary step lengths pseudo-residuals of the PLM-HSMM with $\lambda=(10^{5},10^{4},10^{2})$ which was selected for the muskox movement data. The top left panel shows the histogram of the pseudo-residuals overlaid with the standard normal density function. The upper right panel displays the qq-plot comparing the quantiles of the empirical pseudo-residual distribution and the standard normal distribution. On the bottom, the left panel shows the empirical autocorrelation function and the right panel a sequence of the calculated ordinary pseudo residuals.}
    \label{fig:pr}
\end{figure}

For model checking, we consider ordinary pseudo-residuals as described in \citet{zuc16}. We focus on the pseudo-residuals for the step length observations, as due to their cyclic nature, a residual analysis for turning angles is less amenable. A good model fit is indicated by standard normally distributed pseudo-residuals. Figure \ref{fig:pr} displays the histogram, qq-plot, autocorrelation function, and a time series sequence of the ordinary pseudo-residuals corresponding to the 3-state PML-HSMM with $\lambda=(10^{5},10^{4},10^{2})$ as selected via cross-validation (see main manuscript, Section 3). Overall, the model provides a reasonable fit to the data. 

\renewcommand\refname{References}
\makeatletter
\renewcommand\@biblabel[1]{}

\end{spacing}